# Information overload and environmental degradation: learning from H.A. Simon and W. Wenders

Tommaso Luzzati*†     Ilaria Tucci^     Pietro Guarnieri*

*Department of Economics and Management, University of Pisa, Italy
^TAPRI, University of Tampere, Finland
†*Corresponding author*: tommaso.luzzati@unipi.it

**ABSTRACT**

This paper discusses the relevance of information overload for explaining environmental degradation. Our argument goes that information overload and detachment from nature, caused by energy abundance, have made individuals unaware of the unsustainable effects of their choices and lifestyles. First, the paper shows that concern for information overload brings surprisingly close the works of two outstanding intellectuals engaged in two separated fields, namely the film director Wim Wenders and the social scientist, actually polymath, Herbert Simon. Then, after reviewing the concept of information overload and its influence on decision-making, we show that our argument both can be illustrated within standard consumer theory, and that is consistent with environmental behaviour studies. At the same time interdisciplinary approaches, as ecological economics, are better suited to study the problem. Our research contributes also to show how wide is the scope of application of Simon's ideas, whose relevance to ecological economics has long been recognised.

HIGHLIGHTS:

- Information overload (IO) is a key notion both in Simon's and Wenders' works
- IO and distance from nature lower awareness of environmental side effects of choice
- Low awareness and knowledge help explain environmental degradation

# 1 Introduction

Economists explain environmental degradation with wrong economic incentives. Since the seminal contribution by Pigou (1920), the notion of "externalities" has become the key conceptual tool, and the remedy is taxation to internalize them. The Chicago school approach to the issue, started by Coase (1960), has shifted the focus on the lack of well-defined property rights. This, however, has not changed the narrative (Klink 1994): economists argue that the environment is "consumed too much" because its price is zero (Pearce et al. 1989). While wrong incentives are undoubtedly relevant, the role of other factors remains rather unexplored within economics. This paper discusses the role of awareness of the environmental consequences of actions and lifestyles. Because both time and attention are needed to process information within choice processes, information overload can strongly reduce awareness of the choice consequences. Understanding this phenomenon can prove useful to interpret not only many present-days puzzles like fake-news or beliefs, but also environmental degradation, particularly if one considers the detachment from nature that many people experience today.

Concern for information overload, which was so central for Herbert Simon, who dedicated his life to the study of decision-making, also shaped the (early) works of the film director Wim Wenders. With its transdisciplinary focus, Ecological Economics can gain also from insights outside academia, particularly if they come from highly influential sectors in contemporary life, such as cinema. (Mayumi et al., 2005, is an interesting example). The similarity between Simon and Wenders is highly suggestive both of how close intellectuals working in different fields can be, and how usefully their ideas can be applied to yet another field, environmental behaviour studies. After elaborating on their contributions, in sections 2 and 3, we briefly focus on the concept of information overload and the main findings of the several disciplines that has been dealing with it (section 4). The implications of information overload and energy abundance for environmental degradation are then outlined, also with the help of a standard consumer's choice problem (section 5). In section 6 we discuss our proposal in the light of the behavioural environmental literature, while in section 7 we draw some more general conclusions.

# 2 Wim Wenders

Wim Wenders is a film director, playwright, author, photographer and producer. He has said and wrote a lot about himself, his poetics and way of working, in pamphlets, interviews and published books and there are also many publications about his work. In this section, we will focus on some founding elements of his cinema, particularly his early one. Born in 1945 in Düsseldorf, after a short experience as student first in medicine and then in philosophy, in 1966 he decided to leave Germany for France, where he studied painting and worked as engraver. He was fascinated by the *Cinématheque*, where he developed his passion for movies and in particular for the western genre and begun to write articles and critics for cinema magazines. Back to Germany in 1967, he attended the "Hochschule für Fernsehen und Film München" (University of Television and Film Munich) and made his first short films, in collaboration with other German artists.

Wenders' painting experience strongly influenced his early style and poetic (1970-1984). He considers films as a sequence of images, pictures playing a central role in the composition of the work. Vast images, resembling landscapes' pictures, and slow tracking shots go along the characters through their adventures. This approach also derives from his political and moral conception about cinema, which should allow freely the spectator to think and participate in the dramaturgy construction. In 1971, referring to his films *Summer in the City* and *The Goalkeeper's Fear of the Penalty*, Wenders writes:

> "There are films where you can't discover anything, where there's nothing to be discovered, because everything in them is completely unambiguous and obvious. Everything is presented exactly the way it's supposed to be understood. And then there are other films, where you're continually noticing little details, films that leave the room for all kinds of possibilities. Those are mostly films where the images don't come complete with their interpretations." (Wenders 1991:3)

The German director prefers the second type of films; he believes that films should be cognitive experiences for the spectators, giving them the possibility to freely think, imagine, and create a history in their minds, during the film experience itself. Cinema, in Wenders' opinion, has the ethical responsibility to make spectators-listeners able to decide, to choose which sense or emotion put into the narrative. To this purpose, Wenders uses long times, with vague frames and slow dialogues, along with a balanced and wise rhythm in the editing phase.

The story has not to be violently imposed to the audience by means of visual constraints and forcing images. For the same reason, Wenders has criticized most of television productions, characterized by crowded images, pressed and quick editing, that do not permit to breath and think freely. According to Wenders, the consequences of an overload vision of images are not only on the physio-psychological level of creativity, but they are also moral, with social and political dimensions. Spectators need a complete, definite image, usable in all parts, giving the possibility to move the eyes on the screen with the agility of sensitive and selective activity, without tensions, and art has the responsibility to stimulate the capabilities and the opportunities for a freethinking.

The reason why Wenders wants to involve spectators in the creative process grounds on his conception of filmmaking. He wrote that "films that have a soul" are born out of a dream, or from an intuition, and their aim is to develop it (Wenders, 1997: 18 and ff.) . For example in his work *Until the End of the World*, "the conventionally perceived relation between the visible and the expressive" is turned upside-down, with the consequence that "the expressive (i.e., discursive) is clearly visible and the visible is unutterable" (Kaiser and Leventhal, 2003:142). The film is a puzzle to be solved, and making a movie is a sort of research experience, based on those escaping impressions we can perceive with our peripheral eyes, impressions that are under our eyes, but that we do not embrace and decode completely. Those impressions/sensations peripherally perceived with the corner of the eyes, allow the spectators to think, to freely imagine during the film screening. This is how spectators not only can feel co-author of the artistic work – because they need to put together the elements of dramaturgy to understand what is happening on the screen – but they might feel also a refreshing and regenerating sensation in their minds, due to the imaginative and creative process in which they are involved.

In the ancient Greek culture, thinking and seeing are intrinsically and tightly related. The two terms *ειδοσ* and *ιδεα*, the generic translation of which is "idea", were already in use in the pre-platonic language to indicate *the visible form of things*, what is physically seen, perceived by the eyes. Later on, in his *Dialogs*, Plato uses *ειδοσ* and *ιδεα* with a different understanding; they pertain to the inner form of things, they are the specific objects of the thinking, the true absolute being (Reale, 1988:74). What we can see with the eyes of our bodies are physical things, while what we can see with the eyes of our mind are the non-physical things; the eye of intelligence can see intelligible forms, which are pure essences. It is not by chance that Plato coined the expressions "the mind eye" and "the soul eye" (Reale, 1988:77). Thus, *idea* is not only the essence of the observed things, but also the relation between the object of observation and the observer, something that can be seen by the eyes and the mind. The connection between the act of seeing and the thinking, linguistically and conceptually lies on the ancient root *ιδ,* from which also derives the verbal form of *οιδα,* which literally means "I know because I saw". The etymological Greek intuition that seeing and thinking are strictly related is actually not too far from the effective functioning of physio-psychological system of vision (see, for example, Le Doux, 2003). Eye and mind work together, and they influence each other: the first offers materials to the mind and activates the information storage, the other stimulates the reception abilities/skills of the eye until the fine grade of vision that we normally know/use in our daily life. Therefore, the quality of perception of the external reality – also from the other sensorial channels – can influence our mental processes and consequently our choices.

The connection between seeing and thinking/understanding is at the core of Wittgenstein's reflection on "aspect seeing" or "seeing as"(Wittgenstein, 1953, 1982), i.e. the peculiar cognitive mode that let us grasp the duck or the rabbit in the famous picture popularized by *Gestalt psychology*. Wittgenstein underlines that seeing one (or the other) aspect of such ambiguous images is *per se* (that is, without the mediation of concepts) understanding the meaning of the image and that this understanding consists in a reorganization of the perception such that new relations among the features of the object emerge (Johnston, 2002). These two characteristics of aspect seeing - its immediacy and its relational nature - provide us with further theoretical elements to characterize knowledge as something in between perceiving and thinking. Such a characterization, as we will see, conceptually bridges Simon and Wenders.

Wenders based his cinema on the awareness that the mind reacts very differently, depending on whether the information contained in the visual experience is scarce or excessive. The German director laid great emphasis on the issue of the rate at which the eyes of spectators are stimulated:

> "If there is too much to see, that is, if an image is too full, or if there are too many images, the effect is: you don't see anything anymore. 'Too much' turns quickly into 'nothing'. You all know that. You also know the other effect: if an image is empty, or almost empty, and sparse, it can reveal so much that it completely fills you, and the emptiness becomes 'everything'. […] In a movie you can experience something similar. Some films are like closed walls: there is not a single gap between its images that would allow you to see anything else than what this movie shows you. Your eyes and your mind are not allowed to wander. You cannot *add* anything from yourself to that particular film, no feelings, no experience. You stumble out empty afterwards, like you have been abused. Only those films with gaps in between their imagery are telling stories, that is my conviction. A story only exists and comes to life in the mind of the viewer or listener." (Wenders, 1992: 98-99, original emphasis)

Actually, the ability to process the stimuli (images/information) we are subject to is the necessary premise for rational choice. We process the stimuli, build consistent pictures of our world and figure out possible outcomes of our choices. The relationship between mental capabilities and the environment is at the basis of Simon's contribution, as we will recall in the next section. To acknowledge that minds have some limits in comparison to the decisional environment implies the need of focusing on how humans actually process information when choosing. Simon, then, overturned the mainstream economics concern for scarcity of information, pointing out that the truly scarce resource is the time that an individual is able and available to spend on processing information, that is, mental attention. In a famous quote, he stated:

> "in an information-rich world, the wealth of information means a dearth of something else: a scarcity of whatever it is that information consumes. What information consumes is rather obvious: it consumes the attention of its recipients. Hence a wealth of information creates a poverty of attention and a need to allocate attention efficiently among the overabundance of information sources that might consume it." (Simon, 1971: 40-41)

Both for Simon and Wenders, an excess of information will hamper the ability to think, attention has a key role in understanding and knowing, and emotions affect attention.

## 3   Herbert Simon

As is well-known, Herbert A. Simon (Milwaukee, 1916 – Pittsburgh, 2001) was one of the most influential social scientists of the twentieth century, an eclectic scholar who gave seminal contributions to different disciplines, namely information technology, artificial intelligence, cognitive psychology, and economics. Because of his achievements, he got many important awards, among which, in 1975 (with A. Newell), the most prestigious recognition in computer science, the *Turing Award*, and in 1978 the *Nobel prize for economics*. This section on Simon, whose contributions are very well known, serves not only the purpose of allowing a comparison with Wim Wenders, or highlighting some concepts useful to our argument, but also to show that his key concept of "bounded rationality" cannot be interpreted either as constrained maximisation or imperfect rationality.

The common thread running through his almost thousand publications is the desire to understand decision-making processes, "I am a monomaniac. What I am a monomaniac about is decision-making" he confessed to one of his students (Feigenbaum, 2001: 2107). Interested in knowing how human beings face and solve problems, he spent his scholar life to understand the principles that human mind follows in processing and using information.

In many of his writings, and also in the Nobel Prize lecture (Simon, 1978b), he criticised the notion of rationality used in economics for not considering the limits of human mind. He pointed out many failures and epistemological weaknesses of the neoclassical economic theory, which is grounded on a narrow notion of rationality that does not help explaining empirical observations better than the common sense of rationality. He highlighted that many deductions of neoclassical models do not require the hypothesis of perfect rationality; for instance - as admitted also by a champion of mainstream economics, the Nobel laureate G. Becker - utility maximization is not necessary for obtaining decreasing demand function (Simon, 1978b: 347-9). By bringing numerous examples, Simon showed (e.g. Simon, 1986: S213-215; Simon, 1978a: 4-5) how the deductions of neoclassical theory are grounded more on auxiliary hypotheses (e.g. particular specifications of the model), rather than on an "omniscient" rationality, a term already used in 1959 (Simon, 1959: 265). Incidentally, it has to be recalled that also the Nobel laureate K.J. Arrow, another scholar who contributed to the neoclassical theory, expressed similar concerns. Arrow admitted not only that rationality is often a non- necessary hypothesis (Arrow, 1990: 26), but also that rationality

> "[...] is most plausible under very ideal conditions. When these conditions cease to hold, the rationality assumptions become strained and possibly even self–contradictory. They certainly imply an ability at information processing and calculation that is far beyond the feasible and that cannot well be justified as the result of learning and adaptation." (Arrow, 1990: 25)

The core of Simon´s thesis was the need of attributing a "procedural" meaning to rationality. The mainstream "substantive rationality" is focused only on the results, "[b]ehaviour is substantively rational when it is appropriate to the achievement of given goals within the limits imposed by given conditions and constraints" (Simon, 1976: 66). Neoclassical economics, differently from other social sciences, is not interested in "the nature and origins of values and their changes with time and experience" as well in describing and explaining "the ways in which nonrational processes (e.g., motivations, emotions, and sensory stimuli) influence the focus of attention and the definition of the situation that set the factual givens for the rational processes" (Simon, 1986: S210). When, for example, "psychologists use the term 'rational', it is usually procedural rationality they have in mind", and a "behaviour is procedurally rationality when it is the outcome of appropriate deliberation" (Simon, 1976: 67).

Moving from the choice outcomes to the process, the nature of the problem to be solved becomes crucial; substantive rationality is appropriate and effective for simple problems, while it is not if the decisional context is difficult. For instance, substantive rationality would require playing chess by writing down the entire game tree, i.e., all the possible moves. The reason why nobody, not even computer programs, plays by writing the whole game tree is time (and probably also lack of physical space!):

> "Normally, when a chess player is trying to select his next move, he is faced with an exponential explosion of alternatives. For example, suppose he considers only ten moves for the current position; each of these moves in turn breeds ten new moves, and so on. Searching to a depth of six plies (three moves by White and three by Black) will already have generated a search space with a million paths." (Simon and Chase, 1973: 394).

A player who would try to play chess by writing the game's tree could not be considered as rational. Rather, a rational person is able to adapt the decisional procedure to the difficulties of the task in relation to his/her own computational capacities. The study of the cognitive processes has shown how in the real-life situations, the "difficult" problems are solved by selectively reducing the number of the possible paths. In a similar way operational research tackles the integer programming problems (Simon, 1978a: 11). Humans, who are not abstractly omniscient beings with infinite computational capabilities, explore only a little part of the possible alternatives, addressing only the most promising ones; they follow "rules of thumb" or "heuristics rules" that derive from the identification of patterns and/or trials and errors that come from past experiences (Simon, 1978b: 362).

To cut down the choice domain requires giving up the idea of looking for maximizing solutions but aiming at satisficing solutions, as human beings do in their real lives. Procedurally rational individuals look for "satisficing models that provide good enough decisions with reasonable costs of computation. By giving up optimization, a richer set of properties of the real world can be retained in the models." (Simon, 1978b: 350). At the same time, Simon admitted that, in some instances a maximising solution of a simple model can be "satisficing" in the real world (Simon, 1978b: 350). In any case, 'selective search' and 'satisficing' is a binomial, which defines truly rational behaviour in real decisional contexts, namely procedural rationality (e.g. Simon, 1978b: 356).

Clearly, the attention to the real decisional processes open the doors to other radical changes in the representation of the individual and implies the need to include contributions from cognitive psychology. For example, Simon notes that the level of "aspiration" for which the individual ends his/her own valuation process and considers him/herself satisfied changes depending on how much the surrounding environment is favourable or not (Simon, 1978b: 356). Moreover, individuals acknowledge that ends are not given, but they should be accepted in their adaptation in relation to the means, with the consequence that the individual values change with time and experience (Simon, 1986: S210). Eventually, by focusing on reality, individuals become aware that they are affected by emotions and sensorial stimulations (Simon, 1986: S210).

Unfortunately, the acknowledgement of the complex and dynamic nature of rationality led to a common misunderstanding in behavioural economics, namely the lack of distinction between procedural rationality and "irrationality". Probably, this confusion has been also nurtured by the fact that Simon has coined and used the expression "bounded rationality". This expression, if decontextualized from Simon's thought, is ambiguous and allows two different misinterpretations. On the one hand, Simon's perspective can be reduced to mainstream rational choice, by viewing it as a maximization of given ends under constraints and limits that have not considered before. Even within the information overload literature (see below), sometimes the focus is sometimes on outcomes, that is, on "the decision maker's ability to optimally determine the best possible decision" (Roetzel, 2018: 6). On the other hand, "bounded rationality" can be interpreted as a partially flawed rationality, a mix of rationality and biases, a direction that has been successfully pursued by behavioural economics. Actually, in many of his writings, Simon pointed out that individuals deviate from the standard assumption of neoclassical economics, also by showing irrationalities and biases. However, his bounded rationality was neither constrained optimization, nor imperfect rationality; for him, what is restricted is not rationality but individual computational capacities as compared to the requirements of many real-life choices; as a consequence, the "procedural" one is the highest form of rationality.

The ideas of Simon are relevant within ecological economics both via procedural rationality and hierarchical complexity (Foxon, 2006). Procedural rationality has greatly influenced environmental behavioural economics (see below) and counts as an epistemological foundation for social multicriteria evaluation (Munda 2004), but is also related to the notion of Post-Normal Science (PNS) (e.g. Funtowicz and Ravetz, 1994), which is a constituent of Ecological Economics (see e.g. Castro e Silva and Teixeira, 2011). Consistently with Simon's approach, PNS shows that, when complexity of the decisional context gives rise to high uncertainty and ignorance, science cannot go for truth, but must be guided by quality.

## 4  Information overload

The idea that information can be too much and overburden people is an old one (e.g. Edmunds and Morris, 2000: 19-20, Blair, 2011: 1-2). It is even indirectly referred already in the Ecclesiastes[1], as Bawden and Robinson (2009: 183) stated. Information overload in its modern sense is often traced back (e.g. Klapp, 1986: 98-99) to the sociologist George Simmel [1903, (1950: 413-414)], who argued that city inhabitants are sensorially overburdened by the urban context that makes them jaded and also incapable to react to new situations.

Most probably, the first use of the term "information overload" came from Bertram Gross, who argued that an excess of information available to a person aiming to complete a task or take a decision can negatively influence the process itself and result in a poor (or even no) decision (Gross, 1964:856). During the years, because of its relevance in our contemporary times, this phenomenon has been referred to, studied and discussed in different contexts and disciplines, such as psychology, information technology, health, mass communication, management related disciplines, etc. Moreover, the phenomenon has been changing over the time (see e.g. Bawden and Robinson, 2009). As a consequence, there is not a single definition for it, and different terms are used to refer to the phenomenon, e.g., 'cognitive load', 'data smog', 'information fatigue', 'document tsunami' (Eppler and Mengis, 2004: 326; Roetzel, 2018: 6-7). However, the widely accepted meaning of the term relates to individual's efficiency and accuracy in decision making processes. Information overload occurs when the excessive availability of relevant information becomes an obstacle for processing it and making decisions (Bawden and Robinson, 2009: 182-3). As highlighted by Epple and Mangis (2004: 331), "there is wide consensus today that heavy information load can affect the performance of an individual negatively (whether measured in terms of accuracy or speed)".

---

[1] "of making many books there is no end; and much is study is a weariness of the flesh" Ecclesiastes, Chapter 12, v. 6

The seminal contributions about information overload came from psychology and cognitive science, namely the famous Miller's article "The magical number seven plus or minus two" (Miller, 1956), and two books, respectively by Schroder et al. (1967) and Simon and Newell (1971) (see also Simon, 1979). The developments of the studies on information overload have been summarized by several reviews, most of which have a disciplinary focus, while Bawden et al. (1999), Epple and Mangis (2004), Goetzel (2018) encompass several disciplines.

The general reason why information overload occurs is because the processing requirements exceed the processing capacities. Requirements and capacities are both quantitative and qualitative. Time is an example of a quantitative variable. The time required to process information has to be compared with the time available to individuals (or their willingness to spend time). Qualitative facets relate to personal capacities, to information characteristics (ambiguity, uncertainty, intensity/complexity, novelty, consistency, redundancy), and to characteristics of the task/decision. Moreover, there is an interaction between requirements and capacities since too demanding requirements impair the capacity and the motivation of individuals making them stressed, confused, anxious. As found by the academic literature (see e.g., Epple and Mangis, 2004), individuals react by allocating less time to each information input, ignoring a large part of information by filtering it out, and relying on external sources that synthesise it. Also, identifying the relationship between details and the overall perspective becomes more difficult. The individuals end up needing more time to reach a decision, which results in accuracy loss or even in not taking a decision.

Research that has been inquiring the issue of information overload has focused on several different issues. For instance, Goetzel (2018) organized his reviews by suggesting five categories, namely, the starting situation (e.g task complexity, environment, personal characteristics), the role of the source of information (e.g., information system, database, social media, …), the information search and information processing (e.g. the information characteristics such as its novelty, and conditions of time pressure/restrictions), the subjective informational stance of the decision-maker, and behaviour and emotions after decision-making. To the purpose of the present paper, however, the detailed results of the academic literature are not needed; it is sufficient highlighting that a well-established finding has emerged, namely, that the relationship between the amount of available information and the quality of the decision-making process has an inverted-U shaped curve, as suggested by Schroeder et al (1967) more than 50 years ago. When information is low, its increase will improve the decision-making process. However, beyond some threshold, information becomes not only useless, that is, not included in the process, but also harmful because it will overburden and confuse individuals.

## 5  Implications for environmental degradation

### 5.1  The general argument

In this section we will discuss the consequences of information overload, both in general terms and within the standard consumer's choice between leisure and consumption. According to economics environmental degradation arises because pollution is a public bad and externalities do not enter the individual choice problem. When considering the choice between leisure time and consumption in the presence of negative externality, preferences of a generic individual $j$ can be represented by the following utility function[2]

$$U^j = f(C^j; L^j; P(C^j + C^{-j}))$$

where $C$ is the bundle of consumption goods, $L$ is leisure, and $P$ is pollution.

However, since the side effect of pollution is not immediately observable by the individual and difficult to assess, we add a term that stands for the degree of information accuracy about pollution (which can be interpreted also as awareness) that the individual has in estimating the negative effects that consumption provokes via pollution. By indicating accuracy/awareness with $a(.)$, the utility function becomes

$$U^j = g(C^j; L^j; a(.)P(C^j + C^{-j}))$$

The individual allocates her/his time considering the opportunity cost of leisure time, that is, wage. Normalizing the price of consumption to 1, and assuming that income comes only from labour ($w$

---

[2] For the sake of simplicity only total consumption (and not leisure) is assumed to generate pollution, namely, the sum individual $j$'s and the other individuals' consumption.

is the hourly wage rate), the standard budget constraint will hold, namely,
$$C^j = w(24 - L^j)$$

Because pollution, i.e. a public bad, affects utility, the self-concerned maximising consumption (given the level of pollution generated by the others) will be lower than that in the absence of externalities, but still higher than the efficient bundle, namely the one for which the Samuelson's conditions in the presence of public goods hold (and negative externalities are considered by each individual). Figure 1 traces the three bundles of goods along the budget line, while indifference curves are omitted for a better visualization. The highest bundle is the maximising one in the absence of awareness, which formally coincides with the case of no externalities. The lower the parameter $a(\cdot)$, the lower is leisure and higher are consumption and working time. The intermediate bundle is the maximising bundle that takes into account externalities[3], while the lowest is the efficient bundle (the Lindhal solution), which, as well-known is inconsistent with individual maximization.

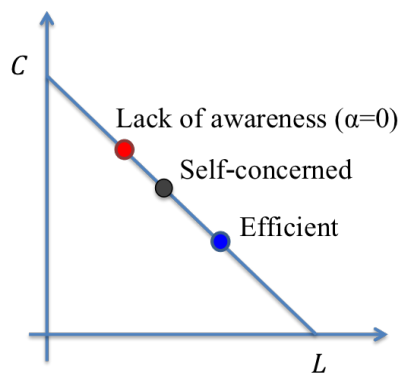

Figure 1. Optimal choices under different hypothesis concerning individual awareness and others' behaviour

5.2  Awareness, information overload, and connection with nature

If one accepts the argument spelled above, the optimal (self-concerned) consumption is determined by the size of the parameter $a$, that is, by awareness/accuracy. What does affect it? An obvious answer is the information available to the individual, $I$. The literature on information overload, as mentioned in the previous section, has shown that the relationship between degree of accuracy and information is inverted-U shaped. Another crucial variable is, as we will try to argue in what follows, is energy use.

People living in very artificial environments, far detached from nature, might face cognitive difficulties in realizing and being aware of the systemic interconnectedness. Indeed, people love nature, enjoy its recreational services, and watch beautiful documentaries; however, their perception is mainly indirect because nature is not part of daily experience. As Martinez Alier put it, "most citizens of the rich urbanised world get their provision from the shops. Hence the proverbial response of urban children to the question of where does the milk or do the eggs come from – the supermarket." (Martinez Alier, 2002, 26). As a result, despite the availability of information, we do not actually see. Hence by not perceiving, we lack full awareness of the consequences of our actions. The change in human lifestyles has been made possible by the exponential increase in the use of fossil fuels that has started with the industrial revolution. Unprecedented levels of exosomatic energy have detached us from nature, providing us comforts that were unconceivable even in a recent past (Georgescu Roegen, 1971). Because energy is invisible its role is difficult to be actually grasped; yet we have become more independent from the external conditions, which implies lower connectedness and hence lower awareness of the systemic consequences of our actions. On the ground of these arguments, we model the degree accuracy/awareness as inversely related to the consumption of exosomatic energy, $E$.

---

[3] Consumption in the "self-concerned" bundle might be lower than in the "efficient" bundle only if the the contribution to reduce pollution by the others is very low. This situation, however, cannot be a Nash equilibrium, which involves pollution higher than the Lindahl solution ("efficient" bundle).

To sum up, we write accuracy/awareness as a function of both available information and exosomatic energy, that is as

$a(I, E)$

with $\frac{\partial a}{\partial E} < 0$; $\frac{\partial^2 a}{\partial E^2} < 0$

and $\frac{\partial a}{\partial I} > 0 \ \forall I < I°, \frac{\partial a}{\partial I} = 0 \ if \ I = I°, \frac{\partial a}{\partial I} < 0 \ \forall I > I°; \ \frac{\partial^2 a}{\partial I^2} < 0$

For illustrative purposes, one can consider the following simple specification

$$a = \frac{I(1 - I)}{\ln(1 + E) + 1/4}$$

which implies that its maximum is $a=1$ (when $I=1/2$ and $E=0$)

By considering that many countries have experienced growth both in information and exosomatic energy consumption, the implication at the macro-level can be easily drawn, namely, that awareness has followed an inverted-U time pattern. When information was low and connection with nature high, more information was highly beneficial; then, progressive detachedness from nature and "artificialization" of our lives have started to off-set the contribute of more information, which eventually ended up to be even detrimental to awareness because of overload. Figure 2 graphically sketches this argument. Perhaps, the strength of the environmental movements at the end of the 1960s corresponded to the peak in the awareness trend drawn in Figure 2; information about the interaction humans-nature was rapidly growing in a context where the detachment from nature was relatively low.

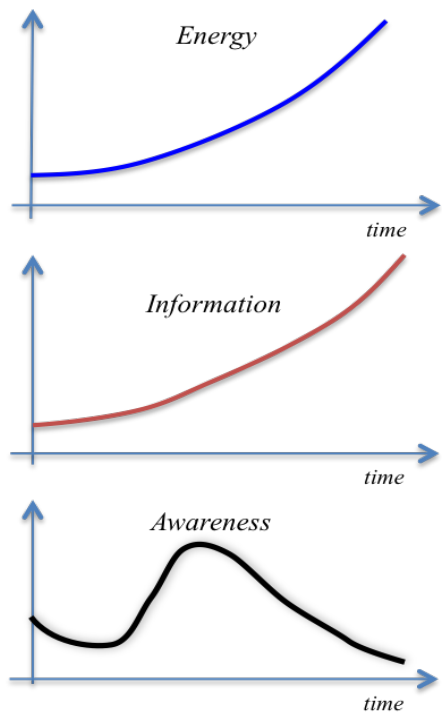

Figure 2. Hypothetical time pattern of awareness as the resultant of trends in information and in exosomatic energy consumption

### 5. 2 Taking into account knowledge

A step further would be considering the distinction between information and knowledge, a distinction that cannot be accommodated within the model of the omniscient homo oeconomicus where, once an agent has complete information, he/she possesses the complete knowledge of the decision-making situation he/she is facing. The elaboration process that connects the acquisition of mere information data to the building of the kind of explanation that constitutes knowledge remains out of the picture.

However, the distinction between information and knowledge has been widely discussed in cognition sciences, philosophy of science and cybernetics. To clarify it, we briefly recall the definition of the two concepts provided within the framework of the *data-information-knowledge-wisdom* theory (DKIW)[4]. The first formulation of this theory is attributed to Kenneth Boulding who distinguished "signals, messages, information, and knowledge" in his 1955 article "Notes on the Information Concept" (Boulding, 1955). The notions included in this theory are object of ongoing re-discussion and clarification (for an overview of the debate see Zins, 2007).

To the purposes of this paper, we refer to the systematization of the theory provided by

---

[4] For an historical introduction see Wallace (2007). For a critical discussion of the developments of the theory see Rowley (2007).

Russell Ackoff – an American systems theorist and professor of organizational change (Ackoff, 1989). This theory is hierarchical since information depends on data, knowledge depends on information, understanding depends on knowledge, and wisdom depends on understanding[5]. Within this framework, data convey mere factual evidence and consists of "symbols that represent the properties of objects and events". In other words, data are discrete and not interrelated pieces of evidence. Data alone do not possess a meaning before it is elaborated into information, that is, "processed data, the processing directed at increasing its usefulness". Hence, information establishes relational connection among data to provide the agent with useful description concerning the "who", "what", "when", "where", and "how many" of a fact or event. The reference to usefulness makes the functional difference between data and information explicit, in the sense that the difference between data and information is given in terms the different function they serve. As far as the difference between information and knowledge is concerned, Ackoff mentions again a functional difference by stating that "knowledge is conveyed by instructions, answers to how-to questions". In other words, knowledge organizes and structures information to serves the practical purposes (even before grasping a complete understanding of phenomena in terms of their "why"). According to Ahsan & Shah (2007) "when information is given meaning by interpreting it, information becomes knowledge. At this point, facts exist within a mental structure that consciousness can process, for example, to predict future consequences, or to make inferences".

Since information needs time to be processed and "converted" into knowledge[6], the choice between consumption and leisure time should also include the (leisure) time spent in processing information[7], $L_K$, while the "pure" leisure time can be indicated as $L_L$. In general, one can assume that "knowledge" contributes the individual utility both directly and indirectly, by improving the awareness of the choice consequences, as written in the following utility function,

$$U^j = h(C^j; L_L^j; a(K, E) P(C^j + C^{-j}); K(I; L_K))$$

while the budget constraint becomes $C^j = w(24 - L_L^j - L_K^j)$.

This, however, while better modelling information overload, would not change qualitatively the outcome; more information and more detachment from nature distort the agent's optimal choice towards levels of consumption and working time higher than those she/he would have chosen if she/he was more aware of the negative externalities of consumption.

---

[5] The discussion of the notions of "understanding" – which Ackoff introduced in the DKIW paradigm – and "wisdom" is out of the scope of this article. For the interested reader, we only report Ackoff's definitions Ackoff refers to understanding as an "appreciation of 'why'", and wisdom as "evaluated understanding". For a presentation of Ackoff's conception see Ahsan & Shah 2007.

[6] The knowledge production function can be thought as having decreasing marginal returns in time and be subject to information overload, for example, $K=I(1-I)L_K^b$, with b<1.

[7] Of course time is needed also to consume, as in Steedman, 2001.

## 6 Discussion: information overload in environmental behaviour studies

The frame set above falls within the behavioural paradigm used in ecological economics and environmental studies, which both emphasize the assumption of bounded rationality (Van den Bergh et al., 2000; Venkatachalam, 2008) and clarify the role of knowledge in pro-environmental decision-making processes (Gkargkavouzi et al., 2019; Dewulf, 2020). Despite various aspects of the relationship between information overload and pro-environmental decisions are dealt with in the literature, these are not organized in a unified theoretical framework and not fully investigated neither empirically, nor experimentally. The conception of knowledge as structured and organised information oriented to practical purposes and its relation with consciousness is implicit in key notions of the Value-Beliefs-Norms theory (Stern and Dietz 1994, Stern 2000) and of the Theory of Planned Behaviour (Ajzen 1991, 2002) – the two main theoretical frameworks through which pro-environmental behaviours are studied[8]. Specifically, Ackoff's notion of "knowledge" (and not "information" or "understanding") seems suitable to a) interpret the "awareness of consequences" grounding norm activation according to Value-Beliefs- Norms theory and b) accounting for with the epistemic foundation of the "Perceived Behavioral control".

At the same time, information overload has been studied in relation to several aspects of consumers' decisions. The literature refers to information overload as a cause of value-attitude gap (McKercher and Prideaux, 2011; Wells et al., 2011; Berthoû, 2013). Indeed, there is evidence that subjects provided with more information feel less personally responsible for, and less concerned about global warming than less informed ones (Kellstedt, Zharan, and Vedlitz, 2008). Information overload may also negatively affect the attention capability of consumers, as is in the case of ecolabels Thøgersen, J. (2000). Thøgersen, J. (2000). The capability of elaborating information in proper knowledge plays a key role in the case of environmental issues whose complex nature makes realizing the connection of lifestyles and consumption patterns with, for example, climate change difficult to realize (Anable, Lane, and Kelay, 2006). Bergstrom et al. (1990) link information overload to the issue of the evaluation of environmental commodities (in terms of the impacts of information on consumers' willingness to pay), by distinguishing an hypothesis of strong information overload accounting for the "emergence of confused or dysfunctional consumer behavior caused by increased information" - and of weak information overload for which "effects of information on preferences diminish at successively higher information levels (see also Grether, D. M., and Wilde 1983). Van der Werff and Steg (2015) discuss whether general campaigns capable to activate norm-oriented energy-saving behaviour may be better suited to avoid information overload than single-purpose policies targeting specific behaviours. On the same line, Abrahamse et al. (2007) study which information framework is able to convey a reduction in consumers' direct and indirect energy uses.

---

[8] Specifically, Ackoff's notion of "knowledge" (and not "information" or "understanding") seems suitable to a) interpret the "awareness of consequences" grounding norm activation according to Value-Beliefs- Norms theory and b) accounting for with the epistemic foundation of the "Perceived Behavioral control" of the. Such applications are not discussed in the present work and left for further research.

# 7 Conclusion

In our 'fast society', mass media supply 'fast-food information ', which needs to be both simple and attractive. We get then used to simplicity and become less willing and able 'to waste' our time and attention to reflect about our complex environment. Hence, our capability of thinking and processing the information inputs becomes weaker; the feedback is a new demand for simplicity from the mass and social media. Many daily phenomena, such as fake-news or perhaps populisms, can be attributed to this self-reinforcing process. This has dramatic consequences in the case of environmental degradation since our direct experience of the natural environment has decreased both in frequency and intensity. The lives of the citizens of the economically advanced countries have never ceased to become more and more artificial. As a result, despite (or because) the increasing information that is available to us, we face difficulties in seeing, perceiving and assessing the consequences of our choices and actions.

Thus, as Wim Wenders emphasised, when the capacity of self-reflection and self-elaboration becomes weaker freedom is at risk: when humans are not autonomous in their capacity of thinking and deciding, they are at risk of manipulation. According to Wenders, for instance

> "American television exploits and stimulates perceptive capacities to the full, thus ending up subjugating them to the schemes of social convention and economic convenience. In this context, "seeing" is no longer an active form of selection and perception: in front of the TV screen there is no longer physical and psychological space and time to develop one's own interior image, one's own position, one's own point of view.. […] Under conditions of effort, pressure, constriction or even violence, the eyes and the mind work poorly, sight is blurred, the conscience becomes manipulable. By contrast, under conditions of relaxation and mobility one can see and think in a natural, personal and clear fashion." (Russo 1997, 51 and 62, our translation)

In this paper, we have highlighted how close are Wim Wenders's and Herbert Simon's ideas. The German movie director has adopted an aesthetics aimed at reducing the information overload. This emerges not only from his movies but also from his writings, where he has explicitly argued against crowded and quick scenes, which result in information overload that the spectator is unable to process (e.g.. Wenders 1986, 1991, and 1997). On the other hand, by going back to Herbert Simon's "procedural rationality" - a term which captures Simon's ideas better than the fuzzy "bounded rationality" - let us highlight the role of the quality of the decision process.

After reviewing the main findings of the academic literature on information overload, we argued that environmental degradation is attributable not only to unpaid externalities but also to weak awareness about the consequences of choices and actions and that this idea can be framed even within a standard consumer problem. At the same time, interdisciplinary approaches, as ecological economics, are better suited to study the issue. In conclusion, we hope to have contributed in showing that information overload and increasing detachedness from nature play a relevant role in the process of environmental degradation. Under this perspective, the framework we proposed might pave the way to further theoretical and empirical research on the role of information quality and knowledge processes in fostering pro-environmental decisions.

**Acknowledgments**